\documentclass[a4paper]{spie}  

\usepackage[
    left=1.93cm,
    right=1.93cm,
    top=2.54cm,
    bottom=4.94cm
]{geometry}
\usepackage{amsmath,amsfonts,amssymb}
\usepackage{graphicx}
\usepackage{xcolor}
\usepackage[colorlinks=true, allcolors=blue]{hyperref}

\title{Sustainability as a design parameter in the early development of the Wide-field Spectroscopic Telescope}

\author[a]{Laurane Fréour}
\author[b]{Etienne Burtin}
\author[c]{Roland Bacon}
\author[a]{Bodo Ziegler}
\author[e]{Sofia Randich}
\author[c]{Arlette Pecontal}
\author[d]{Peter M.\ Weilbacher}
\author[c]{Corentin Cudennec}
\author[f]{Dimitri Buffat}
\author[e]{Simone D'Auria}
\author[g]{Alessandro Meoli}
\author[h]{Letizia P. Cassarà}
\author[c]{Philippe Dierickx}
\author[g]{Vincenzo Mainieri}
\author[h]{Paolo Franzetti}
\author[i]{David Lee}
\affil[a]{Department of Astrophysics, University of Vienna}
\affil[b]{Université Paris-Saclay - CEA - IRFU}
\affil[c]{Centre de Recherche Astrophysique de Lyon }
\affil[d]{Leibniz-Institut für Astrophysik Potsdam}
\affil[e]{INAF-Osservatorio Astrofisico di Arcetri}
\affil[f]{Aix-Marseille Université, CNRS, CNES, LAM UMR7326, Marseille France}
\affil[g]{European Southern Observatory}
\affil[h]{INAF - IASF Milano}
\affil[i]{STFC - UK Astronomy Technology Centre}

\authorinfo{Corresponding author: Laurane Fréour (laurane.freour@univie.ac.at)}

\begin{document} 
\maketitle

\begin{abstract}
The proposed Wide-field Spectroscopic Telescope, a next-generation telescope facility with a 12-meter primary mirror that may be operational in the 2040s, is integrating sustainable considerations from the early design stage. Our preliminary analysis identifies two primary sources of environmental impact that can be mitigated during the design process. 
First, we are conducting a life-cycle assessment (LCA) to quantify the environmental impact of the $\sim$250 spectrographs and the detector cooling systems across the telescope's three instruments, from construction through operation. The carbon footprint estimated through this LCA has been included as a trade-off parameter in the design choices.
Second, the WST is expected to produce 1-3 PB/year, requiring careful evaluation of data processing and storage strategies. We explore approaches to monitoring and reducing the carbon footprint associated with data management.
Bringing environmental factors into the design process from the start gives us the chance to weigh sustainability alongside scientific performance and cost. By treating the carbon footprint as a core design consideration, we aim to make progress toward more responsible astronomy.
 
\end{abstract}

\keywords{Astronomy, Sustainability, Telescope}

\section{INTRODUCTION}
\label{sec:intro}  
Astronomers have a unique perspective on the Earth, its fragility, and the absence of a ‘Planet B’. Yet, their carbon footprint remains substantial. Research infrastructures alone account for about 36 tonnes of CO$_2$ per astronomer per year [\citenum{Knodlseder2022}], nearly 18 times the Paris Agreement’s 2050 target of 2 tonnes of CO$_2$ per person per year. As the climate crisis accelerates, the global temperature rise and the increased likeliness of extreme weather events pose direct threat to astronomy. For instance, recurrent temperatures in excess of the cooling system's abilities challenge the well-functioning of ground-based telescopes [\citenum{2020Cantalloube}], while wildfires can directly threaten the infrastructures. Given the long timescales involved in developing new research infrastructures, incorporating environmental sustainability into their designs is crucial.

A few studies have reported on the environmental impact of scientific instruments (e.g., Athena X-ray integral field unit [\citenum{Barret2024}], ELT instrument MOSAIC [\citenum{Janssen2024}]) or of facilities (e.g., GRAND experiment [\citenum{VARGASIBANEZ2024102903}], Cherenkov Telescope Array mid-sized telescope [\citenum{dosSantosIlha2024}]). These investigations have provided important insights, but they usually examine existing facilities or instruments already in development. As a result, environmental impact is often considered mainly as an outcome to be measured, rather than as a factor that can be minimised by early design decisions. Because the development of new research infrastructure spans many years, bringing sustainability considerations at the very beginning presents a real opportunity to reduce environmental impact in meaningful ways.

The Wide-field Spectroscopic Telescope (WST) is a 12-meter facility currently under consideration [\citenum{Bacon2024},\citenum{Dierickx2024}]. It will be equipped with an integral field unit (IFS) and low-resolution (LR) and high-resolution (HR) multi-object spectrographs (MOS), enabling a large field of view and hundreds of thousands of spectral measurements of galaxies and stars [\citenum{2024whitepaper}].

In this article, we describe how sustainability has been included in the early design of the WST. While sustainability is multi-faceted, relying on three interconnected pillars: society, economy, and the environment, we focus this study on environmental sustainability, in particular using the carbon footprint. We report on two areas where environmental sustainability has been included: as a parameter in the trade-off study of the instruments, presented in Sect.~\ref{sec:cf_tradeoff}, and in the discussions around the data processing, described in Sect.~\ref{sec:data}.

\section{THE CARBON FOOTPRINT AS A TRADE OFF PARAMETER IN THE DESIGN OF THE INSTRUMENTS}
\label{sec:cf_tradeoff}
\subsection{Method description of the carbon footprint computation}
We used a life cycle assessment (LCA) to estimate the environmental impact of the spectrographs and cooling systems for different design options. The process is iterative and structured into four main phases [\citenum{iso14040_2006}]: Goal and Scope Definition, Inventory Analysis, Impact Assessment, and Interpretation. The system boundaries for the LCA are established in the Goal and Scope Definition phase. Data, such as material composition, mass, volume of components, and energy consumption are collected in the Inventory Analysis phase. The Impact Assessment phase translates this raw inventory data into measurable environmental impacts across categories such as climate change, resource use, human toxicity, ecosystem damage, and more. We used the SimaPro software\footnote{\url{https://simapro.com/} - Version 10.2.0.1} coupled with the EcoInvent database\footnote{\url{https://ecoinvent.org/database/}} to link each input (e.g., kg of steel, kWh of electricity) to its associated environmental impact. The analysis can cover a broad range of impact categories, including Climate change (kg CO$_2$-equivalent emissions),  Resource depletion (metal, fossil, and water resources), Human toxicity (impact of hazardous substances on human health), Ecosystem quality (effects on biodiversity and land use), Water scarcity and eutrophication (effects on aquatic systems). However, we will focus on Climate change, providing an estimate of the carbon footprint. We used the ReCiPe assessment methods [\citenum{Huijbregts2017}].
Finally, the Interpretation phase involves analysing the results in relation to the initial goals and scope, identifying the main environmental impacts, and assessing any limitations or uncertainties.

\subsection{Functional unit and system boundaries}
The functional unit is the construction and 1-year operation of one instrument (IFS or MOS-LR or MOS-HR) delivering 3600 hours of observation at nominal performance. This is a comparative LCA aimed at quantifying the environmental impact of several design choices for the instrument, including two detector types (CMOS versus CCD) and associated cooling systems (Linear Pulse Tube versus natural refrigerant relying on CO$_2$).
This study is conducted at a very early stage of the project, and we focus on a few steps where mitigation strategies could be identified. The following steps are included in the LCA: 
\begin{itemize}
    \item Acquisition of raw materials;
    \item Manufacturing processes;
    \item Use of the product.
\end{itemize}

We do not include the transportation of components and materials, the test phase, waste generated, or human labour. We also exclude any end-of-life processes.

The different design names, number of spectrographs, and cooling system types used in this analysis are presented in Table~\ref{tab:spectrographs}. The IFS spectrographs all have two detectors, while the MOS spectrographs have four detectors in both LR and HR cases. The type of detectors considered for each design is indicated next to the name. The type of cooling system and readout electronics is fully determined by the detector type, and is described in Sect.~\ref{subsec:construction}.
Different design names refer to different optical design possibilities and spectrograph configurations.

For the IFS, 13 optical designs are considered [\citenum{Cudennec2026}]. The spectrographs have two spectral arms with a dichroic positioned just before the gratings, and most of them feature dioptric cameras, except for one (B) with a catadioptric design. They are composed of mirrors and lenses made of silica, calcium fluoride, and PBL35Y. The main differences between these designs lie in the number of spectrographs and the size and number of optical components. Generally, designs with fewer spectrographs require more and larger components; conversely, designs with more spectrographs tend to be simpler and more compact. Regarding the 192-spectrograph options specifically, their layouts are very similar, with distinctions arising from the presence or absence of curved detectors and the detector size, which ranges from 60 mm to 90 mm per side.

For HR spectrographs, two different optical designs are currently under study. The first (8M16D) employs an on-axis collimator folded by a flat mirror that accommodates the pseudo-slit [\citenum{Dauria2026}]. Three dichroics create four different spectral channels. Each channel is equipped with a Volume Phase Holographic Grating at a 45° angle of incidence and a dioptric camera. Eight such modules are required to reach the target performance. In the second design, three spectral channels with a pseudo-slit and a collimator are arranged in different geometries and at different scales. This results in a smaller overall module envelope, while the number of modules increases to 16.

Regarding the MOS-LR, we have two similar optical designs derived from the same base: a four-arm spectrograph using 9cm square detectors with 15µm pixels [\citenum{Buffat2026}]. The major difference lies in the field lens. In the 4x9, the field lens is made of fused silica, a material commonly used in optical systems. For the 4x9Y, the field lens is made of Yttrium Aluminium Garnet, a crystalline solid typically doped for use as a gain medium in lasers. Here, the use of this material allows us to achieve better image quality while reducing the asphericity of the lenses in the camera.

\begin{table}[h]
\centering
\caption{Design names, number of spectrographs, number of detectors and cooling systems for the different designs considered in the comparative LCA.\\}
\label{tab:spectrographs}
\begin{tabular}{|l|c|c|l|}
\hline
\textbf{Design Name} & \textbf{Number of Spectrographs}& \textbf{Number of detectors} & \textbf{Cooling System} \\ \hline
\multicolumn{3}{|c|}{\textbf{IFS}} \\ \hline
A - CCD & 144 & 288 & LPT \\ \hline
A - CMOS & 144 & 288 & CO$_2$ \\ \hline
B - CCD & 144 & 288& LPT \\ \hline
B - CMOS & 144 & 288 & CO$_2$ \\ \hline
C - CCD & 128 &256 & LPT \\ \hline
C - CMOS & 128 & 256 & CO$_2$ \\ \hline
D - CCD & 144 & 288 & LPT \\ \hline
D - CMOS & 144 & 288&CO$_2$ \\ \hline
E - CCD & 192 & 384 &LPT \\ \hline
E - CMOS & 192 & 384 &CO$_2$ \\ \hline
F,G - CCD & 192 & 384 &LPT \\ \hline
F,G - CMOS & 192 &384 & CO$_2$ \\ \hline
H - CCD & 192 & 384 &LPT \\ \hline
H - CMOS & 192 & 384 &CO$_2$ \\ \hline
I,J - CCD & 192 & 384 &LPT \\ \hline
I,J - CMOS & 192 & 384 &CO$_2$ \\ \hline
K - CCD & 192 & 384 &LPT \\ \hline
K - CMOS & 192 &384 & CO$_2$ \\ \hline
\multicolumn{3}{|c|}{\textbf{MOS - LR}} \\ \hline
4x9 - CCD & 54 & 216 &LPT \\ \hline
4x9 - CMOS & 54 &216 & CO$_2$ \\ \hline
4x9Y - CCD & 54 & 216 &LPT \\ \hline
4x9Y - CMOS & 54 &216 & CO$_2$ \\ \hline
\multicolumn{3}{|c|}{\textbf{MOS - HR}} \\ \hline
8M16D - CCD & 8 & 32&LPT \\ \hline
8M16D - CMOS & 8 & 32& CO$_2$ \\ \hline
16M4D - CCD & 16 & 64&LPT \\ \hline
16M4D - CMOS & 16 & 64&CO$_2$ \\ \hline
\end{tabular}
\end{table}

\subsection{Construction inputs}
\label{subsec:construction}
The inputs are divided into three main categories: the cooling system, the readout electronics, and the optomechanics.
The Linear Pulse Tube (LPT) cooling system option is based on the Cryocooler LPT9310-HP from Thales Cryogenics. This cryocooler has been used in previous astronomical instruments, such as to cool the detectors of the Dark Energy Spectroscopic Instrument [DESI, \citenum{2016DESI}]. We consider that one LPT is needed per detector.
The CO2-based cooling system is inspired by the technique developed to cool the detectors of several CERN experiments [\citenum{ZWALINSKI2023}]. In that case, we assume that the cooling infrastructure is shared among all detectors of the three instruments. The share of the CO$_2$ cooling system’s construction footprint is distributed proportionally to the number of spectrographs in each instrument. For the IFS, the 128-spectrograph option receives a smaller share than the 192-spectrograph option because it has fewer spectrographs, which reduces cooling demand. The MOS is included in the total spectrograph count, with an estimated 66 spectrographs (midpoint between 62 and 70), ensuring the allocation reflects the cooling system's actual usage intensity. However, as the lowest reachable temperature is -50°C (223K), this system needs to be coupled to another system to reach a suitable temperature for the detectors. In our case, we selected Peltier Thermoelectric coolers.
The read-out electronics considered for the CCDs is the New Generator Controller (NGC) II model [\citenum{Richerzhagen2024}]. Similar to MUSE, we assume that one detector front-end is tied to six detectors [\citenum{Reiss2012}]. For the CMOS, we consider Field-Programmable Gate Arrays (FPGAs), with one FPGA per detector. 

For completeness, we also include a vacuum system. We use a setup similar to DESI's, relying on ion pumps. Each ion pump is tied to a detector and works continuously. At this stage, we have preliminary estimates of the mechanical components (optical bench, enclosure, and base) of the spectrographs only for the MOS HR designs. In that case, we include them in the construction phase.

When possible, we used processes available in SimaPro. For the glasses, we defined tailored processes based on their composition to account for variations among the types. 

We excluded detector construction from the analysis because no data on processing energy were available. 

\subsection{Operation inputs}
For the operation, we consider only the energy consumption of the various systems described in Sect.~\ref{subsec:construction}. 
We assume that the CCDs need to be cooled at 150K. For the CMOS, we use a best-case scenario in which the detectors can operate at 200K and a worst-case scenario in which the temperature must be 180K.
When possible, we define upper and lower bounds for energy consumption. Then, we perform a Monte Carlo analysis in SimaPro to model uncertainties and their impact on the final results. 

For the best-case scenario, we assume that the heat load of the CMOS is 2W, and needs to be cooled to 200K.\footnote{Note that this is pushing both the CMOS and the Peltier technology to the edge of what is currently being done. We assume here that these technologies will be further developed and improved in the next 15 years.} The temperature difference between the hot and cold sides is 30°C. In this configuration, the maximum Coefficient of Performance (COP) is approximately 0.6\footnote{See performance versus current plate in \url{https://www.meerstetter.ch/customer-center/compendium/71-peltier-element-efficiency}}. To be on the safe side, we use a COP of 0.5. This results in a Peltier module power consumption of 4W. 
In the worst-case scenario, the CMOS heat load is 4W and must be cooled to 180K. This requires a two-stage Peltier. The temperature difference between the hot and cold sides is 50°C. The system is highly inefficient, and we consider a COP of about 0.08. The power consumption is then 50W. Tables~\ref{tab:ccd_cooling_impact} and \ref{tab:cmos_cooling_impact} report the energy consumption inputs of options with CCDs and CMOS respectively.

The energy consumption has been modelled using the Chilean mix energy available in SimaPro. We discuss this approximation in Sect.~\ref{sec:next_steps}.

\begin{table}[h]
\centering
\small
\caption{Energy consumption associated with different parts for the design options with CCDs cooled using LPT cryocoolers. The energy consumption is scaled per detector.\\}
\label{tab:ccd_cooling_impact}
\begin{tabular}{|p{4.3cm}|c|c|p{6.5cm}|}
\hline
\textbf{Component} & \textbf{Max [W/day]} & \textbf{Min [W/day]} & \textbf{Information} \\
\hline
CCD & 216 & 72 & We assume that the dissipated power from the CCD is equivalent to the energy consumption, varying between 3W and 9W. \\
\hline
LPT & 2160 & 864 &
The LPT is operating at 50\% of the maximum cooling power of 180W in the worst-case scenario and 20\% in the best-case scenario. \\
\hline
Ionic pump & 480 & 480 & The vacuum is maintained by an ion pump for each detector. We consider the continuous operation of the ion pump, with a maximum power consumption of 600W.\\
\hline
Readout electronics - NGC II & 2600 & 1000 & Preliminary computations suggest that one detector front-end consumes between 250W and 650W and is linked to six detectors.\\
\hline
Chiller consumption & 1410 & 686 & We assume that a chiller with a Coefficient of Performance (COP) of 4 removes all the heat generated by the different parts. \\
\hline
\end{tabular}
\end{table}

\begin{table}[h]
\centering
\small
\caption{Energy consumption associated with different parts for the design options with CMOS cooled using CO$_2$ refrigerant and Peltier thermoelectric coolers. The energy consumption is scaled per detector.\\}
\label{tab:cmos_cooling_impact}
\begin{tabular}{|p{4.3cm}|c|c|p{6.5cm}|}
\hline
\textbf{Component} & \textbf{Max [W/day]} & \textbf{Min [W/day]} & \textbf{Information} \\
\hline
CMOS & 96 & 48 &
In the best-case scenario, CMOS dissipate 2W and 4W in the worst-case scenario. \\
\hline
Thermoelectric cooler (Peltier-type) & 1200 & 96 & See main text. \\
\hline
CO2 cooling system - & 451.2 & 177.6 &
We assume a COP of 5. This assumption is likely to be optimistic and will need to be revised once more data is available.\\
\hline
Vacuum system  & 480 & 480 &
We consider a vacuum system similar to the option with LPT. \\
\hline
Readout electronics - HYDRA (FPGA) & 360 & 360 &
The dissipated power is estimated to be 15W per unit. \\
\hline
Chiller consumption & 872.8 & 474.4 &
Assuming a COP of 4. \\
\hline
\end{tabular}
\end{table}

\subsection{Results}

The goal of this analysis was threefold. First, we wanted to compare the carbon footprint of CCD technologies cooled by a mechanical cryocooler (LPT) and of CMOS detectors cooled by a liquid refrigerant (CO$_2$) coupled to a thermoelectric cooling device. Second, we wanted to estimate the share of the carbon footprint attributable to the construction and operation phases. 
Finally, we were also interested in comparing different designs within an instrument to answer whether designs with more spectrographs but lighter have a smaller carbon footprint than designs with fewer spectrographs but heavier.  

   \begin{figure} [ht]
   \begin{center}
   \begin{tabular}{c} 
   \includegraphics[height=7cm]{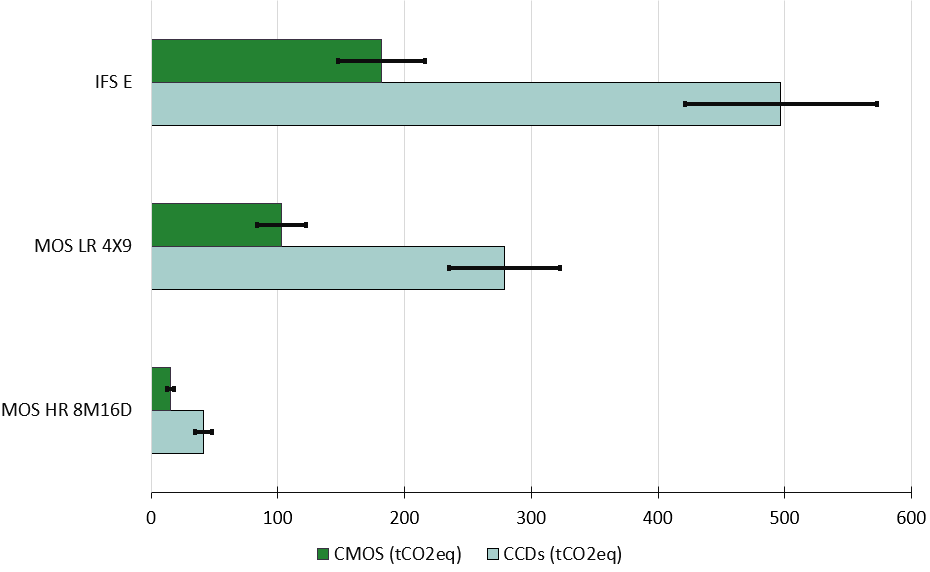}
   \end{tabular}
   \end{center}
   \caption[example] 
   { \label{fig:CMOS_vs_ccd} 
Comparison of the 1-year operation carbon footprint for options with CDDs (blue) and with CMOS (green) for one design of the IFS (upper), MOS LR (middle), and MOS HR (lower).}
   \end{figure} 

Figure~\ref{fig:CMOS_vs_ccd} shows the difference in the one-year operational carbon footprint for the designs with CCDs, in blue, and the designs with CMOS, in green. For all instruments, CMOS options reduce the operational carbon footprint by about 60\%, thanks to lower energy consumption. CMOS can operate at higher temperatures than CCDs, thefore requiring less cooling, and the read-out electronics associated with it, which relies on the FPGA technology, consumes significantly less energy than the NGC II associated with CCDs. The CO$_2$ cooling system is also highly efficient and can be shared between instruments. We note here that, while it would not be possible to use a CO$_2$ based cooling system to cool CCDs due to the temperature threshold, it is possible to use FPGAs with CCDs, or NGCII with CMOS. We did not include these two options in the analysis, but using FPGAs as read-out electronics for the CCDs could decrease the carbon footprint. The MOS HR instrument has the lowest operational carbon footprint because it has the fewest detectors (32 for the 8M16D and 64 for the 16M4D). 

\begin{figure} [ht]
   \begin{center}
   \begin{tabular}{c} 
   \includegraphics[height=9cm]{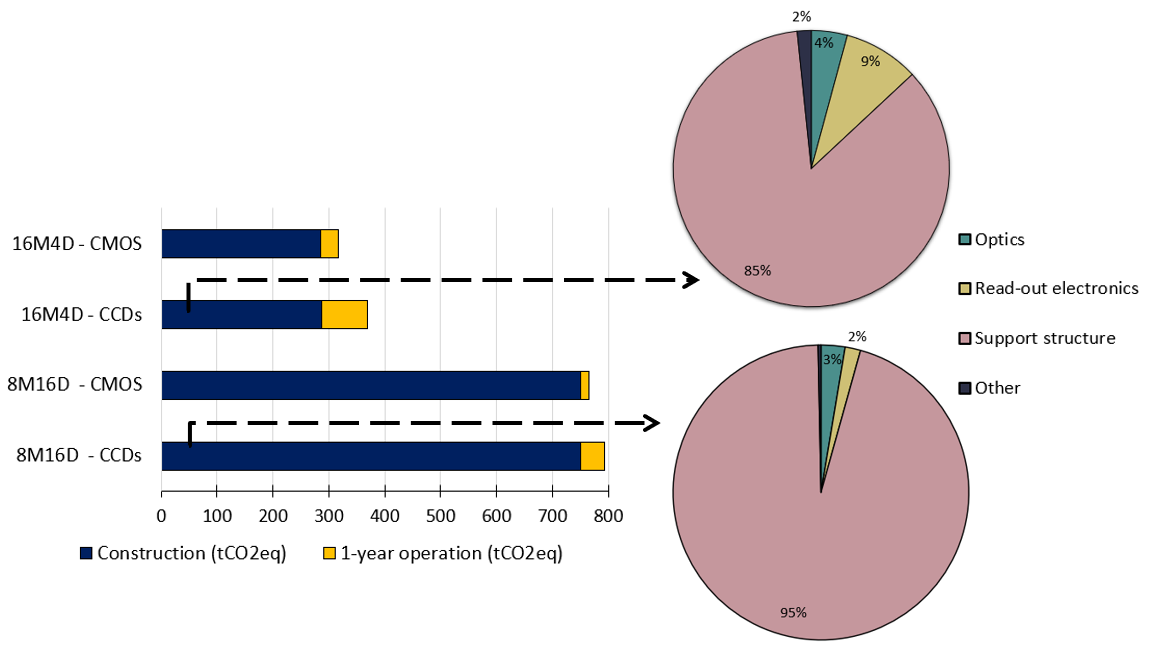}
   \end{tabular}
   \end{center}
   \caption[example] 
   { \label{fig:MOS_HR} 
Comparison of the construction and 1-year operation carbon footprints for the MOS HR designs, with details on the distribution of the carbon footprint across the different components in the construction for the design 16M4D and 8M16D with CCDs.}
   \end{figure}

To assess the second point, on the share of the carbon footprint attributable to the construction and operation phases, we study the different designs of the MOS HR, as we have the most complete data inventory for this instrument. Figure~\ref{fig:MOS_HR} shows the carbon footprint from construction and 1-year operation for the four different MOS HR designs.

In three of the four designs investigated, the carbon footprint associated with operation dominates that of construction after a few years. For the 16M4D designs, the operational carbon footprint dominates after 9 years for the CMOS option and after 3 years for the CCDs option. For the latter, we show on the right panel of Fig.~\ref{fig:MOS_HR} the share of the carbon footprint of the different components in the construction. The optical bench, enclosure, and base of the spectrographs largely dominate, accounting for 85\% of the total construction footprint. This is due to the high mass of aluminium and the associated manufacturing required for these components. The read-out electronics, often requiring copper and other materials with high environmental impact due to extraction processes, account for the second-largest share of the carbon footprint from construction, 9\%. 
This distribution is similar among other MOS-HR designs, with a share from the support structure of 95\% (design 8M16D). For the 8M16D option with CCDs, the operational carbon footprint exceeds that of construction after ~16 years, whereas it takes more than 44 years for the CMOS option, due to the low energy consumption relative to the large, heavy spectrographs in this design. 

 These results confirm previous studies showing that the operational carbon footprint dominates after a few decades (see for e.g., [\citenum{Knodlseder2022}]). They also highlight the necessity of considering both the construction and operational footprints when designing new instruments and keeping a long-term perspective in mind. Designs that might have the lowest carbon footprint over a year can have a significantly higher footprint after 20 years of operation. For instance, the 16M4D design with CCDs, due to its much smaller construction footprint, performs better on a short timescale than the 8M16D with CCDs. However, after 20 years, the latest has a carbon footprint that is about 400 tCO2eq smaller. We reflected this in the trade-off study by attributing a weight to the carbon footprint aggregated over 20 years.

   \begin{figure} [ht]
   \begin{center}
   \begin{tabular}{c} 
   \includegraphics[height=7cm]{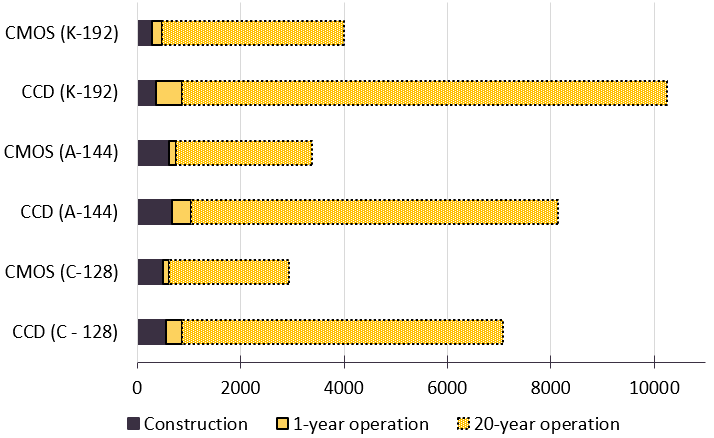}
   \end{tabular}
   \end{center}
   \caption
   { \label{fig:CF_IFS} 
Comparison of the carbon footprint (tCO2eq) for 3 different design options for the IFS, ranging from an option with 128 spectrographs (C), to 144 spectrographs (A), to 192 spectrographs (K) for CCDs and CMOS.}
   \end{figure} 

The last point we investigated tackles the number of spectrographs. In Fig.~\ref{fig:CF_IFS}, we plot the carbon footprint related to the construction and 20-year operation of several design options with a different number of spectrographs, for both CCDs and CMOS options for the IFS. Contrary to the construction of the MOS HR, here we did not include any information on the enclosures and optical benches of the spectrographs, as such information is not yet available. This means the carbon footprint associated with constructing the different designs might be significantly underestimated. However, this does not affect the trade-off study, as the aim here is to compare the carbon footprints of different design options for the IFS.
In all cases, we find that designs with fewer spectrographs have a lower carbon footprint, but this decrease is much smaller than the dominant effect of detector and cooling system choices. Indeed, for a given cooling system, the design C with 256 detectors has a total carbon footprint of 2940 tCO2eq versus 3994 tCO2eq for the design K with 384 detectors. This represents a 26\% decrease, while shifting the difference between using CCDs and LPT and CMOS and CO$_2$ results in a 58\% decrease for design C. 

\section{ENVIRONMENTAL CONSIDERATIONS FOR DATA HANDLING}
\label{sec:data}
The WST aims to be a time-domain and big-data facility. A challenge is managing substantial volumes and complexity in spectra and datacubes while maintaining rapid processing for alert systems and consistent survey results, all with minimal environmental impact. 

In this section, we provide a first estimate of the carbon footprint of several data processing steps: data transfer, data reduction, and data storage.
We explore three different scenarios. In scenario 1, the raw data is transferred every night to Garching, Germany, where one of the European Southern Observatory's offices is located, and reduced and stored there. In scenario 2, the same steps are completed in France, where the carbon intensity of the energy mix is amongst the lowest in Europe. Finally, in scenario 3, the data is stored in Chile, where the telescope would operate. The data is then processed and stored there. 

First, we need to estimate the volume of data generated by the telescope every night. Using a preliminary count of 650 detectors and assuming that data volume scales proportionally with the number of pixels and detectors, we estimate that the WST will generate 1.2 TB of raw data per night and 2.5 TB of data in total (raw, calibrated, and reduced) per night. 

In scenarios 1 and 2, this data is transferred to Europe, most likely by optical fibres. The carbon footprint of the transfer arises mainly from embodied impacts from optical fibres and devices along the way. We estimated it using emission factors from [\citenum{loygue2025}], which drew on several studies from different networks. The computation accounts for the routing distance, optical fibre infrastructure and shelters.
The geographical distance between La Silla and Garching (Germany) is approximately 11,400 km. However, real-world data transfer routes are indirect due to submarine cables and terrestrial detours. We applied a circuitousness ratio of 1.8 [\citenum{Landa2013}] to estimate the routing distance, which is the geographical distance multiplied by the circuitousness ratio.
Shelters are essential for maintaining signal strength over long distances. We assumed one shelter every 100 km along the routing path. The carbon footprint per shelter was reported at $2.47 \times 10^{-5}$ kg CO$_2$eq/GB in [\citenum{loygue2025}], originally computed based on the RENATER network [\citenum{renater2022carbon}]. The carbon footprint of the optical fibre infrastructure was taken as $4.31\times 10^{-3}$ kg CO$_2$eq/GB [\citenum{loygue2025}]. The total carbon footprint for data transfer was computed by multiplying the factors listed above by the amount of data generated.

Then, raw data undergo calibration and reduction across three distinct levels, producing intermediate products and a final data catalogue (e.g., stellar positions and radial velocities). The main carbon footprint of this step comes from the energy consumed by the hardware to reduce the data.
We estimated the IFS needs by extrapolation from the MUSE instrument, after timing a full science reduction [\citenum{Weilbacher2020}] on a modern multi-core machine. We then assume a twice as dense parallelization, and scale the 24 original spectrographs with 4k CCDs to the expected WST setup with 6k detectors, assuming a linear run-time behaviour per pixel. We similarly extrapolate the MOS computations based on the runtime of the L1 pipeline of the 4MOST instrument [\citenum{Worley2022}], scaling the typical runtime for its nine detectors to the WST setup.
The carbon footprint of this process is quantified using the Green Algorithm [\citenum{Lannelongue2021}] to estimate the energy required. Then, we multiplied the energy by the carbon intensity of the energy grid\footnote{Values taken from \url{https://lowcarbonpower.org/map}} in Germany (scenario 1), France (scenario 2), and Chile (scenario 3).

Finally, all raw, calibration, and reduced data must be stored, resulting in a growing data volume over time. The carbon footprint depends on storage type. Here, we used the Solid State Disk from Seagate, model Nytro 3530, and used the energy consumption per GB of 20.7
kg CO2e/TB-yr \footnote{\url{https://www.seagate.com/gb/en/sustainability/planet/product-sustainability/}}. This allows us to multiply by the carbon intensity of the energy grid in the different countries, as in the data reduction step.

\begin{figure} [ht]
   \begin{center}
   \begin{tabular}{c} 
   \includegraphics[height=11cm]{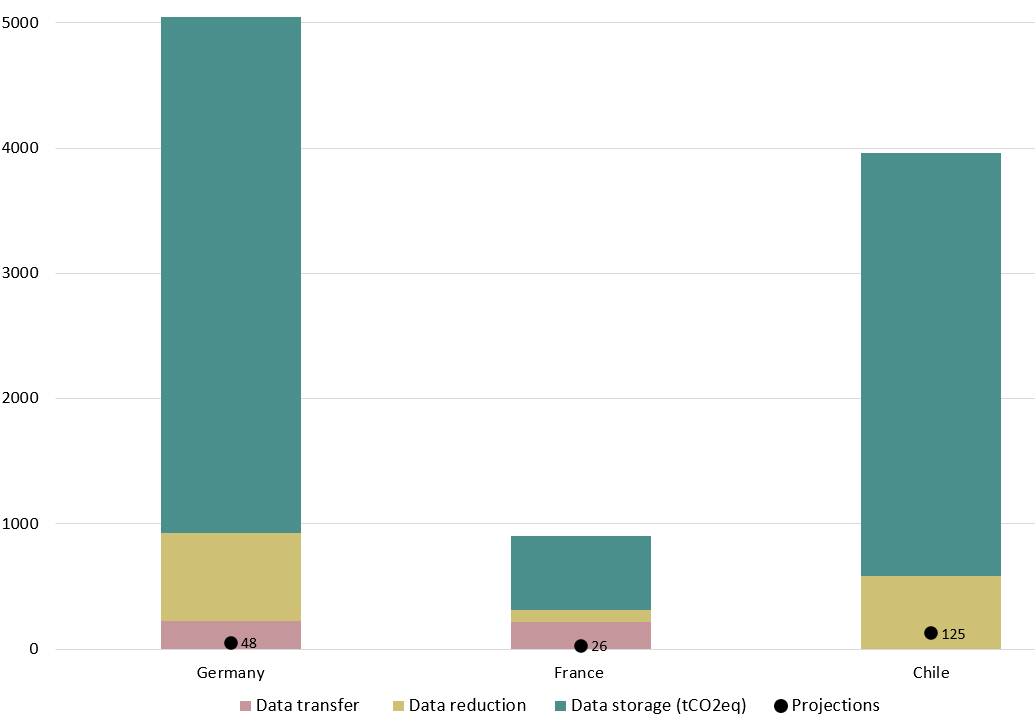}
   \end{tabular}
   \end{center}
   \caption{20-year cumulative carbon footprint associated with different data processing steps for three scenarios: data transferred, reduced and stored in Germany (scenario 1), in France (scenario 2) or kept and processed in Chile (scenario 3). The black dots and corresponding numbers (in tCO2eq) indicate how the footprint could evolve if we consider the carbon intensity forecasted in the 2050s, for both the Data reduction and Data storage steps.}
   \label{fig:data_processing} 

   \end{figure}

The results of the analysis are plotted in Fig.~\ref{fig:data_processing}, which shows the share of the 20-year cumulative carbon footprint for the three scenarios, divided into data transfer (pink), data reduction (gold), and data storage (cyan). Data storage accounts for the largest share of the footprint across all scenarios, due to accumulated data volume over the years. Indeed, we assumed that the data generated each year is added to the data from the previous years. A way to mitigate this is to put some data, the kind that does not need to be accessed regularly, in ``cold" storage. The highest carbon footprint arises for scenario 1, where the data is transferred to Germany, and processed and stored there. The 20-year carbon footprint is 5050 tCO2eq. In comparison, the operational carbon footprint of option K of the IFS (with 384 CMOS detectors) is approximately 4000 tCO2eq (see Fig.\ref{fig:CF_IFS}). In scenario 3, where all the steps take place in Chile, the carbon footprint of data transfer is set to 0. We note that this could change if the system boundary is modified. Indeed, most data users are based in Europe. If the data is stored in Chile, it will still need to travel across the Atlantic for each user request. However, this is currently out of the scope of this study. 

The carbon footprint of data reduction and storage has been computed using the current carbon intensity mix (308.76, 44.12, and 253.11 gCO2eq/kWh for the three scenarios, respectively). These values will surely change in the future, with global trends suggesting a decrease in the carbon intensity of energy worldwide. To quantify this change in the results, we used continental predictions based on countries' Nationally Determined Contributions and announced pledges. The carbon intensity forecasted in the 2050s is 10 gCO$_2$eq/kWh in Europe and 48 gCO$_2$eq/kWh in South America\footnote{\url{https://eneroutlook.enerdata.net/forecast-world-co2-intensity-of-electricity-generation.html}}. In this case, the 20-year carbon footprint of the data reduction and storage drops to 48 tCO$_2$eq in the case of Germany, 26 tCO$_2$eq for France and 125 tCO$_2$eq for Chile. These new values are shown as black dots in Fig.~\ref{fig:data_processing}.

\section{Next steps}
\label{sec:next_steps}
The analyses presented in this article are ongoing. In particular, LCA analysis requires regular updates to reflect changes in the inputs and designs. In this section, we highlight a few steps that need further development or revision. 

First, a few inputs need to be included or revised in the LCA. For example, we did not include the materials and manufacturing of the detectors, nor the mechanical support structure for the MOS-LR and IFS, in the construction phase due to a lack of available data. This leads to an underestimate of the construction footprint. The energy inputs provided in Tables \ref{tab:ccd_cooling_impact} and \ref{tab:cmos_cooling_impact} are initial estimates that will require regular updates as the cooling systems evolve. In particular, we note that the COP of the CO$_2$ cooling system is likely to be lower than the one adopted in this analysis, resulting in higher energy consumption than planned. However, this could be counterbalanced by the fact that such a CO$_2$ cooling system might not require an external chiller to remove the heat. Future refinements will enable more precise energy consumption estimates. 
As mentioned in Sect.~\ref{sec:cf_tradeoff}, using Peltier technology at temperatures as low as 200K is currently a major challenge and may be highly inefficient. To avoid using it, a solution could be to use a krypton refrigerant coupled to the CO$_2$ loop [\citenum{contiero2024krypton}]. This promising solution is currently under investigation.

We would also like to investigate the carbon footprint of other aspects that have not yet been considered. In particular, we identified the dome's cooling as potentially having a high carbon impact. Indeed, the dome needs to be kept at a very stable temperature to avoid thermal distortions in the telescope structure, air turbulence that would degrade image quality, and condensation on optical surfaces that could damage sensitive equipment. This can be particularly energy-intensive, depending on the external temperature. Quantifying and exploring different cooling solutions would allow us to develop the best-suited cooling system. This is all the more important because extreme weather events are very likely to increase as the climate is disrupted, with more frequent droughts and days with temperatures above seasonal means in Chile [\citenum{seneviratne2021weather}]. We would also like to investigate potential energy systems to power the telescope. Previous studies have shown that a combination of solar panels, hybrid energy storage, and a backup diesel generator is both economically and sustainably viable [\citenum{VIOLE2023128570},\citenum{VIOLE2024123334}] for telescopes in Chile. Such studies provide valuable information on energy performance in the next decades. 

Regarding the carbon footprint analysis of data processing, further refinement of our computations is required. In particular, we plan to review and implement storage policies to minimize the volume of data retained and determine which data sets can be transitioned to ``cold" storage. Regarding data reduction, our current assessment considers only one step, whereas there are actually three steps that lead to the data ultimately being made publicly available. As we gather more information about the various pipelines, we intend to include the carbon footprint of all reduction steps in our analysis.

Finally, the carbon footprint is a key indicator of environmental impact, but it does not capture the full picture. The LCA analysis highlighted that other significant environmental impacts, such as ecotoxicity, resource scarcity, and human toxicity, are equally important to consider. For instance, components such as the readout electronics (FPGA) and the Peltier module exhibit high levels of ecotoxicity. This is primarily due to the use of toxic heavy metals and hazardous chemicals in semiconductor manufacturing. Including the full environmental impact in the trade-off analysis would enable us to better capture the different impact categories of the designs.


\section{CONCLUSION}
In this article, we discussed how we integrated environmental sustainability into the early design of the WST. Using an LCA method,  we calculated the carbon footprint of 26 different designs for the IFS and 4 different designs for the MOS HR and LR. In particular, we analysed the potential reduction in carbon emissions between designs using CCD detectors, mechanical cryocoolers (type LPT), and NGCII read-out electronics, and designs using CMOS detectors with a CO$_2$ cooling system and FPGAs. We find that designs relying on CMOS and CO$_2$ cooling systems have a $ \sim$60\% decrease in annual operational footprint compared with options using CCDs and mechanical cryocoolers. The distribution between the construction footprint and the operational footprint depends heavily on the mass and number of spectrographs and on the energy consumption. Designs with lighter spectrographs but more numerous usually have a lower carbon footprint during construction due to lighter optics and support structures. However, these designs have a larger operational footprint due to more detectors to cool and more readout electronics, resulting in higher energy consumption. While it is easier to mitigate the carbon footprint from the operational phase, shifting to renewable energies, for example, mitigating the carbon footprint from the construction phase might be more complicated, requiring exploring sustainable procurement and the use of recycled materials. 
The results from the carbon footprint computation have been included in the trade-off study conducted to select the best design for the IFS and MOS instruments, alongside risk, cost, and scientific performance.

We also analysed the carbon footprint of several data processing steps across three scenarios, depending on their locations. The carbon footprint depends highly on the energy mix of the countries where the data processing and storage will take place. In the case of Germany, which has the highest electricity carbon intensity among Chile and France, the 20-year carbon footprint of data transfer, first-reduction step, and storage is estimated at close to 5050 tCO2eq. In comparison, the 20-year operational carbon footprint of the three instruments is estimated at about 6000 tCO2eq, a number subject to variation depending on the final designs selected.
These findings highlight the importance of incorporating sustainability considerations as early as possible, particularly in the design of new scientific instruments and facilities, where meaningful reductions can be achieved. By continuously evaluating and minimizing environmental impacts, we can ensure that future technological developments and astronomical discoveries are achieved responsibly and with a reduced carbon footprint.

\acknowledgments 
 
This project has received funding from the European Union Horizon Europe Research and Innovation Action under grant agreement no. 101183153 -WST. Views and opinions expressed are however those of the author(s) only and do not necessarily reflect those of the European Union or the European Research Executive Agency (REA). Neither the European Union nor the REA can be held responsible for them. LF is grateful to 
Jürgen Knödlseder and Leidy-Tatiana Vargas-Ibanez for the helpful discussions about Life Cycle Assessments. 

\section*{DATA AVAILABILITY}
An Excel table gathering the inputs used for the life cycle inventory, the SimaPro tailored processes, and the results of the analysis will be made public on the WST website\footnote{\url{https://wstelescope.eu/}}, shortly after the publication of this paper.

\bibliography{report} 

@article{Knodlseder2022,
  author = {Knödlseder, J. and Brau-Nogué, S. and Coriat, M. and al.},
  title = {Estimate of the carbon footprint of astronomical research infrastructures},
  journal = {Nature Astronomy},
  volume = {6},
  pages = {503--513},
  year = {2022},
  doi = {10.1038/s41550-022-01612-3},
  url = {https://doi.org/10.1038/s41550-022-01612-3}
}

@article{ZWALINSKI2023,
title = {Progress in new environmental friendly low temperature detector cooling systems development for the ATLAS and CMS experiments},
journal = {Nuclear Instruments and Methods in Physics Research Section A: Accelerators, Spectrometers, Detectors and Associated Equipment},
volume = {1047},
pages = {167688},
year = {2023},
issn = {0168-9002},
doi = {https://doi.org/10.1016/j.nima.2022.167688},
url = {https://www.sciencedirect.com/science/article/pii/S0168900222009809},
author = {L. Zwalinski and P. Barroca and C. Bortolin and al.}
}

@article{Richerzhagen2024,
author = {Mathias Richerzhagen and Matthias Seidel and Leander Mehrgan and al.},
title = {{System design of the newest generation detector controller for extremely large telescope and new very large telescope instruments}},
volume = {10},
journal = {Journal of Astronomical Telescopes, Instruments, and Systems},
number = {4},
publisher = {SPIE},
pages = {046001},
year = {2024},
doi = {10.1117/1.JATIS.10.4.046001},
URL = {https://doi.org/10.1117/1.JATIS.10.4.046001}
}

@ARTICLE{2016DESI,
       author = {{DESI Collaboration} and {Aghamousa}, Amir and {Aguilar}, Jessica and al.},
        title = "{The DESI Experiment Part II: Instrument Design}",
      journal = {arXiv e-prints},
         year = 2016,
        month = oct,
          eid = {arXiv:1611.00037},
        pages = {arXiv:1611.00037},
          doi = {10.48550/arXiv.1611.00037},
archivePrefix = {arXiv},
       eprint = {1611.00037},
 primaryClass = {astro-ph.IM},
       adsurl = {https://ui.adsabs.harvard.edu/abs/2016arXiv161100037D}
}

@article{Reiss2012,
author = {Reiss, Roland and Deiries, Sebastian and Lizon, J.-L and Rupprecht, G.},
year = {2012},
month = {09},
pages = {},
title = {The MUSE instrument detector system},
volume = {8449},
journal = {SPIE},
doi = {10.1117/12.925388}
}

@article{loygue2025,
  TITLE = {{Carbon footprint of cloud, edge, and Internet of Edges}},
  AUTHOR = {Loygue, Pauline and Al Agha, Khaldoun and Pujolle, Guy},
  URL = {https://hal.science/hal-05450979},
  JOURNAL = {{Annals of Telecommunications - annales des t{\'e}l{\'e}communications}},
  PUBLISHER = {{Springer}},
  VOLUME = {80},
  NUMBER = {1},
  PAGES = {153-169},
  YEAR = {2025},
  MONTH = Feb,
  DOI = {10.1007/s12243-024-01061-1},
  HAL_ID = {hal-05450979},
  HAL_VERSION = {v1}
}

@ARTICLE{2020Cantalloube,
       author = {{Cantalloube}, Faustine and {Milli}, Julien and {B{\"o}hm}, Christoph and al.},
        title = "{The impact of climate change on astronomical observations}",
      journal = {Nature Astronomy},
         year = 2020,
        month = sep,
       volume = {4},
        pages = {826-829},
          doi = {10.1038/s41550-020-1203-3},
archivePrefix = {arXiv},
       eprint = {2009.11779},
 primaryClass = {astro-ph.IM},
       adsurl = {https://ui.adsabs.harvard.edu/abs/2020NatAs...4..826C}
}

@article{Lannelongue2021,
author = {Lannelongue, Loïc and Grealey, Jason and Inouye, Michael},
title = {Green Algorithms: Quantifying the Carbon Footprint of Computation},
journal = {Advanced Science},
volume = {8},
number = {12},
pages = {2100707},
doi = {https://doi.org/10.1002/advs.202100707},
url = {https://advanced.onlinelibrary.wiley.com/doi/abs/10.1002/advs.202100707},
eprint = {https://advanced.onlinelibrary.wiley.com/doi/pdf/10.1002/advs.202100707},
year = {2021}
}

@inproceedings{Landa2013,
author = {Landa, Raul and João, Taveira and Araújo, Richard and al.},
year = {2013},
month = {01},
pages = {},
title = {The Large-Scale Geography of Internet Round Trip Times},
journal = {2013 IFIP Networking Conference, IFIP Networking 2013}
}

@report{renater2022carbon,
  author = {Ficher, Marion and Berthoud, Françoise and Ligozat, Anne-Laure and al.},
  title = {Rapport : évaluation de l’empreinte carbone de la transmission d’un Gigaoctet de données sur le réseau {RENATER}},
  year = {2022},
  url = {https://www.renater.fr/wp-content/uploads/2022/01/empreinte_carbone_transport_donnees_renater_compresse.pdf},
  type = {Rapport},
  institution = {Réseau National de Télécommunications pour la Technologie, l'Enseignement et la Recherche (RENATER)}
}

@inproceedings{contiero2024krypton,
  author = {Contiero, L. and Verlaat, B. and Hafner, A. and al.},
  title = {Applying krypton as refrigerant for cooling of future particle detector trackers at {CERN}},
  year = {2024},
  month = {August},
  booktitle = {16th IIR Gustav Lorentzen Conference on Natural Refrigerants (GL2024)},
  address = {College Park, Maryland, USA},
  url = {https://doi.org/10.18462/iir.gl2024.1235}
}

@incollection{seneviratne2021weather,
  author = {Seneviratne, S.I. and Zhang, X. and Adnan, M. and al.},
  title = {Weather and Climate Extreme Events in a Changing Climate},
  booktitle = {Climate Change 2021: The Physical Science Basis. Contribution of Working Group I to the Sixth Assessment Report of the Intergovernmental Panel on Climate Change},
  year = {2021},
  publisher = {Cambridge University Press},
  address = {Cambridge, United Kingdom and New York, NY, USA},
  pages = {1513--1766},
  doi = {10.1017/9781009157896.013}
}

@article{Barret2024,
  author = {Barret, Didier and Albouys, Vincent and Knödlseder, Jürgen and al.},
  title = {Life cycle assessment of the {ATHENA} {X}-ray integral field unit},
  journal = {Experimental Astronomy},
  volume = {57},
  number = {3},
  pages = {19},
  year = {2024},
  month = {04},
  day = {25},
  issn = {1572-9508},
  doi = {10.1007/s10686-024-09939-7},
  url = {https://doi.org/10.1007/s10686-024-09939-7}
}

@article{dosSantosIlha2024,
  author = {dos Santos Ilha, Gabrielle and Boix, Marianne and Knödlseder, Jürgen and al.},
  title = {Assessment of the environmental impacts of the {Cherenkov Telescope Array} mid-sized telescope},
  journal = {Nature Astronomy},
  volume = {8},
  number = {11},
  pages = {1468--1477},
  year = {2024},
  month = {11},
  day = {01},
  issn = {2397-3366},
  doi = {10.1038/s41550-024-02326-4},
  url = {https://doi.org/10.1038/s41550-024-02326-4}
}

@article{VARGASIBANEZ2024102903,
title = {Life cycle analysis of the GRAND experiment},
journal = {Astroparticle Physics},
volume = {155},
pages = {102903},
year = {2024},
issn = {0927-6505},
doi = {https://doi.org/10.1016/j.astropartphys.2023.102903},
url = {https://www.sciencedirect.com/science/article/pii/S0927650523000890},
author = {Leidy T. Vargas-Ibañez and Kumiko Kotera and Odile Blanchard and al.}
}

@inproceedings{Janssen2024,
author = {Annemieke Janssen and Kacem El Hadi and Nazim Ali Bharmal and al.},
title = {{Estimate of the environment impact of the ELT instrument MOSAIC}},
volume = {13099},
booktitle = {Modeling, Systems Engineering, and Project Management for Astronomy XI},
editor = {S{\'e}bastien Elias Egner and Scott Roberts},
organization = {International Society for Optics and Photonics},
publisher = {SPIE},
pages = {130990P},
year = {2024},
doi = {10.1117/12.3018865},
URL = {https://doi.org/10.1117/12.3018865}
}

@inproceedings{Dierickx2024,
author = {P. Dierickx and T. Travouillon and G. Gausachs and al.},
title = {{WST - Widefield Spectroscopic Telescope: design of a new 12m class telescope dedicated to widefield multi-object and integral field spectroscopy}},
volume = {13094},
booktitle = {Ground-based and Airborne Telescopes X},
editor = {Heather K. Marshall and Jason Spyromilio and Tomonori Usuda},
organization = {International Society for Optics and Photonics},
publisher = {SPIE},
pages = {1309431},
year = {2024},
doi = {10.1117/12.3018518},
URL = {https://doi.org/10.1117/12.3018518}
}

@ARTICLE{2024whitepaper,
       author = {{Mainieri}, Vincenzo and {Anderson}, Richard I. and {Brinchmann}, Jarle and al.},
        title = "{The Wide-field Spectroscopic Telescope (WST) Science White Paper}",
      journal = {arXiv e-prints},
         year = 2024,
        month = mar,
          eid = {arXiv:2403.05398},
        pages = {arXiv:2403.05398},
          doi = {10.48550/arXiv.2403.05398},
archivePrefix = {arXiv},
       eprint = {2403.05398},
 primaryClass = {astro-ph.IM},
       adsurl = {https://ui.adsabs.harvard.edu/abs/2024arXiv240305398M}
}

@inproceedings{Bacon2024,
author = {Roland Bacon and Vincenzo Maineiri and Sofia Randich and al.},
title = {{WST - Widefield Spectroscopic Telescope: motivation, science drivers and top level requirements for a new dedicated facility}},
volume = {13094},
booktitle = {Ground-based and Airborne Telescopes X},
editor = {Heather K. Marshall and Jason Spyromilio and Tomonori Usuda},
organization = {International Society for Optics and Photonics},
publisher = {SPIE},
pages = {130941O},
year = {2024},
doi = {10.1117/12.3018093},
URL = {https://doi.org/10.1117/12.3018093}
}

@Article{Weilbacher2020,
  author        = {{Weilbacher}, Peter M. and {Palsa}, Ralf and {Streicher}, Ole and al.},
  title         = {{The Data Processing Pipeline for the MUSE Instrument}},
  journal       = {A\&A},
  year          = {2020},
  volume        = {641},
  pages         = {A28},
  month         = sep,
  adsurl        = {https://ui.adsabs.harvard.edu/abs/2020A&A...641A..28W},
  archiveprefix = {arXiv},
  doi           = {10.1051/0004-6361/202037855},
  eid           = {A28},
  eprint        = {2006.08638},
  primaryclass  = {astro-ph.IM}
}

@INPROCEEDINGS{Worley2022,
       author = {{Worley}, C.~C. and {Walton}, N.~A. and {Murphy}, D.~N.~A. and al.},
        title = "{Rehearsing the complex data flow of multi-object spectrograph survey projects}",
    booktitle = {Modeling, Systems Engineering, and Project Management for Astronomy X},
         year = 2022,
       editor = {{Angeli}, George Z. and {Dierickx}, Philippe},
       series = {Society of Photo-Optical Instrumentation Engineers (SPIE) Conference Series},
       volume = {12187},
        month = aug,
          eid = {1218706},
        pages = {1218706},
          doi = {10.1117/12.2629253},
       adsurl = {https://ui.adsabs.harvard.edu/abs/2022SPIE12187E..06W}
}

@article{Huijbregts2017,
  author = {Huijbregts, Mark A. J. and Steinmann, Zoran J. N. and Elshout, Pieter M. F. and al.},
  title = {ReCiPe2016: a harmonised life cycle impact assessment method at midpoint and endpoint level},
  journal = {The International Journal of Life Cycle Assessment},
  year = {2017},
  month = {02},
  pages = {138--147},
  volume = {22},
  number = {2},
  doi = {10.1007/s11367-016-1246-y},
  url = {https://doi.org/10.1007/s11367-016-1246-y},
  issn = {1614-7502}
}

@INPROCEEDINGS{Cudennec2026,
       author = {{Cudennec}, C. and {Jeanneau}, A. and {Bacon}, R. and al.},
        title = "{Guiding Design Choices for Wide-Field IFS: Trade-Offs
Between Replication and Complexity for WST}",
         year = 2026
}

@INPROCEEDINGS{Dauria2026,
       author = {{D'Auria}, S. and {Tozzi}, A. and {Brucalassi}, A. and al.},
        title = "{WST, the Wide-field Spectroscopic Telescope: Mechanical Design and FE
Analyses for the High Resolution Spectrograph}",
         year = 2026
}

@INPROCEEDINGS{Buffat2026,
       author = {{Buffat}, D. and {Dohlen}, K. and {Saunders}, W. and al.},
        title = "{WST, the Wide-field Spectroscopic Telescope: design trade-offs for the low-resolution multi-object spectrograph instrument}",
         year = 2026
}

@article{VIOLE2023128570,
title = {A renewable power system for an off-grid sustainable telescope fueled by solar power, batteries and green hydrogen},
journal = {Energy},
volume = {282},
pages = {128570},
year = {2023},
issn = {0360-5442},
doi = {https://doi.org/10.1016/j.energy.2023.128570},
url = {https://www.sciencedirect.com/science/article/pii/S0360544223019643},
author = {Isabelle Viole and Guillermo Valenzuela-Venegas and Marianne Zeyringer and Sabrina Sartori},
}

@article{VIOLE2024123334,
title = {Integrated life cycle assessment in off-grid energy system designâ€”Uncovering low hanging fruit for climate mitigation},
journal = {Applied Energy},
volume = {367},
pages = {123334},
year = {2024},
issn = {0306-2619},
doi = {https://doi.org/10.1016/j.apenergy.2024.123334},
url = {https://www.sciencedirect.com/science/article/pii/S0306261924007177},
author = {Isabelle Viole and Guillermo Valenzuela-Venegas and Sabrina Sartori and Marianne Zeyringer}
}

@standard{iso14040_2006,
  author = {{International Organization for Standardization}},
  title = {ISO 14040:2006 -- Environmental management -- Life cycle assessment -- Principles and framework},
  edition = {2},
  year = {2006},
  month = {07},
  date = {2006-07},
  organization = {ISO}
}
\bibliographystyle{spiebib} 

\end{document}